\magnification 1200
\baselineskip 15 pt
\font\meinfont=cmbx10 scaled \magstep3
\font\deinfont=cmbx10 scaled \magstep2
\nopagenumbers
\line{\hfill BI-TP 94/29}
\line{\hfill June 1994}
\vskip 1truecm
\centerline{\meinfont Susceptibilities, the Specific Heat and }
\centerline{\meinfont a Cumulant in Two-Flavour QCD }
\medskip
\vskip 2 truecm
\centerline{\deinfont Frithjof Karsch and Edwin Laermann}\medskip
\centerline{Fakult\"at f\"ur Physik, Universit\"at Bielefeld,
Postfach 10 01 31,}\par
\centerline{D-33501 Bielefeld, Germany}\medskip
\vskip 3 truecm
\centerline{\deinfont Abstract}\medskip
{\narrower
We study the quark mass
dependence of various response functions, which contribute to
chiral
susceptibilities and the specific heat
in the staggered fermion formulation of two-flavour QCD. This
yields information about the critical exponents $\alpha$, $\beta$
and $\delta$. In the case of the chiral susceptibility, obtained as
derivative of the chiral order parameter with respect to the
quark mass, we calculate all contributions. This allows
to construct a cumulant of the order parameter, which is a
scaling function and yields a direct determination of the critical
exponent $\delta$. All our results are consistent with a second order
phase transition.
\par}

\vfill\eject
\pageno=1
\noindent
{\deinfont 1) Introduction}
\medskip
In contrast to QCD with three or more light flavour degrees of
freedom,
where numerical simulations seem to indicate that the finite
temperature chiral symmetry restoring phase transition is first
order,
there is increasing evidence that the two-flavour theory has a
second
order phase transition in the limit of vanishing quark masses. At
least
in the parameter range studied so far in the staggered
formulation of
lattice QCD, no signs of a first order
transition have been observed [1]. In fact, in a recent analysis
[2] of the
existing Monte Carlo data for pseudo-critical couplings,
$g_c(m_q)$,
at non-vanishing quark mass, one of us has shown that the
functional
dependence on $m_q$ is consistent with the scaling behaviour
expected
for a chiral phase transition with critical exponents in the
universality
class of a 3-d $O(4)$ symmetric $\sigma$-model [3,4]. Still, even
in the
class of $O(4)$ symmetric models the occurrence of first order
phase
transitions can not be ruled out easily [3,5]. A careful analysis of
the nature of the chiral transition is thus asked for.

In most of the existing studies of the chiral phase transition in
the staggered fermion formulation of two-flavour QCD [6,7]
the critical parameters have been determined through a subjective
inspection of the dependence of the chiral order parameter or the
Polyakov loop expectation value on the gauge coupling.
In Ref.~[8] the volume and quark mass dependence of the critical
couplings have been examined by locating the maximum of the variance
of these quantities.  They are, however, not directly related to bulk
thermodynamic quantities and predictions for their
scaling behaviour can, therefore, not directly be obtained
from properties of the two-flavour partition function, ie. the
singular part of the corresponding free energy. A priori it thus is not
clear in how far these quantities can give information on the critical
exponents of two-flavour QCD. We will discuss this question here in
detail.

In this paper we will study susceptibilities and the specific
heat obtained from second derivatives of the partition function,
$$\eqalign{
C_V &= {1\over VT^2}{\partial^2 \over \partial (1/T)^2} \ln Z~~, \cr
\chi_m &= {T\over V}{\partial^2 \over \partial m_q^2} \ln Z~~, \cr
\chi_t &= - {T\over V}{\partial^2 \over \partial m_q\partial (1/T)}
\ln Z~~. \cr
}  \eqno(1)
$$
The determination of the maxima of these quantities
will allow us to give a precise quantitative definition of the
pseudo-critical couplings at non-vanishing values of the quark
mass. We
will discuss the scaling behaviour of these couplings as well as
that of the peak heights in the response functions.

From $\chi_m$ and the chiral order parameter
$$
\langle \bar\psi \psi \rangle = {N_F \over 4}{T\over V}{\partial \over
\partial m_q} \ln Z
\eqno(2)
$$
we will construct the cumulant
$$
\Delta  = { m_q \chi_m \over \langle \bar\psi \psi \rangle}~~,
\eqno(3)
$$
which allows a direct investigation of the zero-mass critical coupling
as well as the critical exponent $\delta$ [9]. In such an analysis
it is particularly important to determine the complete chiral
susceptibility, $\chi_m$, ie. taking into account also the
contribution from the connected part in the product of the
quark bilinears.

This paper is organized as follows. In the next section we will
summarize basic relations for the critical behaviour of thermodynamic
quantities in the vicinity of a second order phase transition,
which is controlled by the temperature as well as an external symmetry
breaking field, ie. the quark mass.
In section 3 we discuss the observables from which critical exponents
for QCD with dynamical quarks on the lattice are calculated. In
section 4 our Monte Carlo data for various response functions
is presented for
three values of the quark mass on lattices of size $8^3 \times 4$.
The determination of the critical exponents $\alpha$, $\beta$
and $\delta$ is discussed. Finally we give our Conclusions in
section 5.

\bigskip\bigskip
\noindent
{\deinfont 2) Scaling Relations}
\medskip

If the chiral phase transition in two-flavour QCD is second
order, the critical behaviour of thermodynamic quantities will
be controlled by two external parameters, the reduced temperature
$t=(T-T_c)/T_c$ and the symmetry breaking field $h\equiv m_q/T$.
In the staggered formulation of lattice regularized QCD these two
dimensionless couplings are given by
$$
\eqalign{
t &= {6 \over {g^2}} - {6 \over {g^2_c (0)}}~~, \cr
h &= m_q N_\tau~~ .}\eqno(4)
$$
Here $g^2_c (0)$ denotes the critical coupling on a lattice of
temporal extent $N_\tau$ in the limit of
vanishing bare quark mass. For non-vanishing values of the quark
mass, a pseudo-critical coupling, $g^2_c (m_q)$, can, for instance,
be defined as the location of a peak in one of the susceptibilities
or the specific heat defined in eq.~(1).

In the vicinity of the critical point the behaviour of bulk
thermodynamic quantities is governed by thermal ($y_t$) and
magnetic $(y_h)$ critical exponents, which characterize the
scaling behaviour of the singular part of the free energy density,
$$
f(t,h)\equiv - {T \over V} \ln Z =b^{-1}f(b^{y_t}t, b^{y_h} h)~~.
\eqno(5)
$$
Here $b$ is an arbitrary scale factor. It is expected that the
chiral phase transition in QCD can be described by an effective,
three dimensional theory for the chiral order parameter, which in
the case of two-flavour QCD would amount to an $O(4)$ symmetric
spin model [3,4]. Still the generic structure of this effective
theory leaves open the possibility of a first order transition
[3,5]. A quantitative analysis
of the scaling behaviour of thermodynamic quantities is needed to
further support the existence of a second order phase transition
in the zero quark mass limit.
In the staggered lattice discretization of QCD the situation gets
even more involved due to the fact that
flavour symmetry is partially broken by ${\cal O} (a)$
terms, where $a$ is the lattice spacing. Correspondingly, at
least at large lattice spacings, the relevant symmetry group is
$O(2)$ [9], which would lead to somewhat different critical exponents
and even might influence the order of the phase transition [9].
It remains to be seen if and when the full flavour symmetry is
effectively restored on lattices with small but finite spacings.

The scaling behaviour of the specific heat and susceptibilities
is controlled by the critical exponents $\alpha = {(2y_t-1)/y_t}$,
$\beta =(1-y_h)/y_t$ and
$\delta=y_h/(1-y_h)$. Their numerical values for $O(2)$ and
$O(4)$ symmetric spin models in three dimensions are given in
Table I.

\medskip
$$
\offinterlineskip \tabskip=0pt
\vbox{
\halign to 1.0\hsize
{\strut
 \vrule width0.8pt\quad#
 \tabskip=0pt plus 50pt
 &#\quad
 &\vrule#&
 &\quad #
 &\vrule#
 &\quad #
 &\vrule#
 &\quad #
 &\vrule#
 &\quad #
 &\vrule#
 &\quad #
 &\vrule#
 &\quad #
  \tabskip=0pt
 &\vrule width 0.8pt#\cr
 \noalign{\hrule}\noalign{\hrule}
 & && $\alpha$ && $\beta$ && $\delta$ && $z_\alpha$&& $z_m$&&
$z_t$ & \cr
\noalign{\hrule}\noalign{\hrule}
 &$O(2)$&& -0.007(6) && 0.3455(20) && 4.808(7)&&-0.004(3) &&
0.792(1)&&
0.394(2) & \cr
 &$O(4)$&& -0.19(6) && 0.38(1) && 4.82(5)&&-0.10(3) && 0.793(3)&&

0.34(1) & \cr
\noalign{\hrule\hrule}}}
$$
\medskip
{\narrower
Table I: Critical exponents of 3-d $O(N)$ symmetric spin
models [4,10,11].
The exponents $z_m$ and $z_t$ characterize the scaling behaviour
of susceptibilities as defined in eq.~(6). The exponent $z_\alpha$
is defined in eq.~(10) and describes the leading mass-dependence of
the specific heat peak.
\medskip}

Eq.~(5) can be used to extract scaling laws for various
quantities, valid in the vicinity of the critical point.
For the location and height of the maximum in the
susceptibilities one finds
$$\eqalign{
t_{\rm max} &= c \;m_q^{\,1/\beta \delta}~~; \cr
\chi_{m,{\rm max}} &= c_{m} \;m_q^{-z_m}
= c_{m} \;m_q^{\,1/\delta-1}~~; \cr
\chi_{t,{\rm max}} &= c_{t} \;m_q^{-z_t}
= c_{t} \;m_q^{\,(\beta-1)/\beta\delta}~~.
}\eqno(6)
$$
For finite values of the quark mass the pseudo-critical couplings
defined through peaks in $\chi_m$ or $\chi_t$ can, of course,
differ.
The exponents $z_t$ and $z_m$, which control the scaling
behaviour of
the thermal and chiral susceptibilities are also given in Table I.

Similarly we obtain for the scaling behaviour of the chiral order
parameter on the zero-mass critical point $(t=0)$ as well as on
the line of pseudo-critical couplings $(t=t_{\rm max})$,
$$
\langle \bar\psi \psi \rangle = c_\psi m_q^{1/\delta}~~.
\eqno(7)
$$

In the case of spin models the analysis of cumulant ratios, which
themselves are scaling functions, has proven to be useful. The
simplest
ratio one can consider in the case of QCD is the ratio of the
second and
first derivative of the free energy with respect to the quark
mass, ie. the ratio
$$
\Delta (t,h) = {1\over N_{\tau}}{ h \chi_m \over
\langle \bar\psi \psi \rangle}~~.
\eqno(8)
$$
As can be seen again from eq.~(5) this ratio is a scaling function,
which only depends on the combination $y= th^{-1/\beta\delta}$. A
consequence of this is that $\Delta (t=0, h)$ is unique for all
values of $h$ (of course, modulo corrections from the regular part
of the free energy). Outside the critical region the order
parameter $\langle \bar\psi \psi \rangle$ is expected to depend
linearly on the quark mass,
$\langle \bar\psi \psi \rangle = c_0 + c_1 m_q$ with $c_0$ being
non-zero for $t < 0$ and zero otherwise. From this one obtains
$$
\Delta (t,h) \equiv Q\bigl( th^{-1/\beta\delta}\bigr)=\cases{
1 & $t<0~,~h \rightarrow 0$ \cr
1/\delta & $t=0$ \cr
0 & $t>0~,~h \rightarrow 0$ }~~.
\eqno(9)
$$

Similarly one can analyze the scaling behaviour of the specific
heat peak. Like in the case of the susceptibilities, the quark mass
dependence of its location is given by $t_{\rm max}$ defined in eq.~(6).
However, unlike for the susceptibilities discussed above, the analytic
calculations for the exponent $\alpha$ suggest that the
specific heat does not diverge in the zero quark mass limit ($\alpha < 0$)
but scales as
$$
C_{V,{\rm max}}  = c_0 + c_1 \;m_q^{-z_\alpha} \quad; \quad
z_\alpha = \alpha/\beta\delta~~.
\eqno(10)
$$
The mass dependence thus is a subleading term.

\bigskip\bigskip
\noindent
{\deinfont 3) Determination of Critical Exponents on the Lattice}

\medskip

In the following we will discuss the observables from which
information
on the critical exponents $\alpha$, $\beta$ and $\delta$ is
extracted. These are
various response functions which contribute to the
specific heat and certain susceptibilities defined for two-flavour
QCD. In particular, we will study the behaviour of three
susceptibilities
- the chiral ($\chi_m$) and thermal ($\chi_t$) susceptibilities,
defined in eq.~(1), as well as the Polyakov loop response function
$\chi_L$,
$$
\chi_L = N_{\sigma}^3 \left\{ \langle L^2 \rangle - \langle L
\rangle^2 \right\}
~~,\eqno(11)
$$
where $L= {1\over 3}N_{\sigma}^{-3}\sum_{\vec x}{\rm Tr}
\prod_{i=1}^{N_\tau}
U_{(\vec x,i),\hat 0}$ denotes the average Polyakov loop
on a lattice of size $N_{\sigma}^3 \times N_\tau$.
While the critical behaviour of the first two can be extracted
from the structure of the free energy density, the latter is not
directly related to it nor does it serve as an order parameter
for a
symmetry of the Lagrangian at finite quark mass. In fact, it is
expected
that $\langle L \rangle$ does not show any critical behaviour
(divergence) in the limit of vanishing quark mass. Nonetheless,
the existing
Monte Carlo investigations show the presence of a pronounced
peak in
the Polyakov loop response function, which reflects the sudden
onset of deconfinement in the chirally symmetric phase.

On the lattice the chiral susceptibility $\chi_m$ for $N_F$
flavours is obtained from
matrix elements of the staggered fermion matrix, $D=m_q {\bf 1}
+\sum_{\mu}
D_{\mu}$, as
$$
\chi_m = \chi_0 + \chi_{\rm conn}~~,
\eqno(12)
$$
with
$$\eqalign{
\chi_0 &= {N_F \over 16 N_{\sigma}^3 N_{\tau}} \left\{
\langle\bigl( {\rm Tr} D^{-1}\bigr)^2  \rangle -
\langle {\rm Tr} D^{-1}\rangle^2 \right\} ~~, \cr
\chi_{\rm conn} &= - {N_F \over 4} \sum_x \langle \,D^{-1}(x,0)
D^{-1}(0,x) \,\rangle~~.
} \eqno(13)
$$
The first term, $\chi_0$, gives the fluctuations of the order
parameter
$$
\langle \bar\psi \psi \rangle = {N_F \over 4N_{\sigma}^3 N_{\tau}}
\langle {\rm Tr} D^{-1} \rangle
~~.\eqno(14)
$$
The second term, $\chi_{\rm conn}$, results from the connected
part
appearing in the second derivative of the fermion determinant. It
is
the integral over the connected part of a scalar propagator.

The thermal susceptibility, $\chi_t$, requires to calculate
the derivative of the
chiral condensate with respect to the temperature, which in terms
of the
temporal lattice spacing, $a_{\tau}$, and the number of lattice
points in
that direction, $N_\tau$, is given by $T= (N_\tau a_\tau)^{-1}$.
Following the standard procedure for taking derivatives with
respect to
the temperature on the lattice [12] we obtain
$$\eqalign{
\chi_t &= {1 \over N_\tau} {\partial \over \partial a_\tau}
\langle \bar\psi \psi \rangle \cr
&=N_{\sigma}^3 \left\{\langle \bar\psi \psi \cdot \epsilon
\rangle -
\langle \bar \psi \psi
\rangle \langle \epsilon \rangle \right\}\cr }~~,
\eqno(15)
$$
where $\epsilon$ denotes the energy density operator [12]
$$\eqalign{
\epsilon &=
3 \left[-{6\over g^2}+6{\partial g^{-2}_s \over\partial\xi}\right]
P_{\sigma}
+ 3 \left[ {6\over g^2}+6{\partial g^{-2}_t \over\partial\xi}\right]
P_{\tau} \cr
&+ \bar\psi D_0 \psi
+ {\partial m_q \over \partial \xi} \bar\psi \psi ~~. \cr }
\eqno(16)
$$
Here $P_{\sigma (\tau)} = {1\over 9}N_\sigma^{-3}N_\tau^{-1}
\sum_P{\rm Re~Tr} \prod_{i \in P} U_i$
denotes the usual plaquette operator for spacelike (timelike)
plaquettes; $\bar\psi D_0 \psi = {1\over 4} N_F N_\sigma^{-3}N_\tau^{-1} {\rm
Tr}D_0D^{-1}$ while $\xi=a_\sigma / a_\tau$ is the ratio of spatial and
temporal lattice spacing and the derivatives in eq.~(16) should be
evaluated for $\xi=1$. For further details we refer to Ref.~[12].

The calculation of $\chi_t$ gets complicated due to the fact that
the derivatives of the couplings with respect to the temporal lattice
spacing are only known perturbatively. These perturbative relations,
however, are poor approximations to the exact derivatives in the
coupling regime we are presently investigating. Moreover, for a complete
determination of $\chi_t$ one should also
take into account the zero temperature part of the energy
density, which still has to be subtracted in eq.~(16). Fortunately,
all this is not essential for the discussion of the scaling
behaviour of $\chi_t$, ie. its divergence in the limit of vanishing
quark mass. The zero temperature terms will not show any singular
behaviour at the finite temperature phase transition point and
the derivatives appearing in eq.~(16) can be treated as constant
factors. They will not influence the singular behaviour.
We thus can study separately the scaling behaviour of the
different terms contributing to $\chi_t$,
$$\eqalign{
\chi_{t,\sigma} &= \langle \bar\psi \psi \cdot P_\sigma \rangle -
\langle \bar \psi \psi \rangle \langle P_\sigma \rangle  \cr
\chi_{t,\tau} &= \langle \bar\psi \psi \cdot P_\tau \rangle -
\langle \bar \psi \psi \rangle \langle P_\tau \rangle  \cr
\chi_{t,f} &= \langle \bar\psi \psi \cdot \bar\psi D_0 \psi
\rangle -
\langle \bar \psi \psi \rangle \langle \bar\psi D_0 \psi \rangle
\cr
}\eqno(17)
$$
Like in the case of the chiral susceptibility, the last response
function, $\chi_{t,f}$, requires the calculation of contributions from
connected diagrams. These have been omitted in the
following analysis. Our results for $\chi_m$ seem to justify this.
Recall that, in contrast to $\chi_m$, we are here only interested
in the divergent part of $\chi_t$. For the same reason we also omit the
contribution coming from the correlation of $\bar\psi \psi$ with the mass
term in eq.~(16). Due to the additional mass factor this part will not
contribute to the singular behaviour of $\chi_t$.
Each of the other partial contributions to the thermal
susceptibilities may diverge in the zero quark mass limit.
In general, they even may diverge with different critical exponents,
$$
\chi_{t,x} \sim m_q^{-z_{t,x}}~~.
\eqno(18)
$$
The critical exponent, $z_t=(1-\beta)/\beta\delta$, which
controls
the divergence of $\chi_t$ in the zero quark mass limit, is then
given
by the maximum of these three exponents
$$
z_t = {\rm max}\{z_{t,\sigma},~z_{t,\tau},~z_{t,f}\}~~.
\eqno(19)
$$

Similarly we can investigate the scaling behaviour of the
specific heat, defined in eq.~(1). We can ignore all
non-singular contributions and constant factors and
simply investigate the scaling behaviour of the various response
functions contributing to $C_V$;
$$\eqalign{
C_{V,\mu\nu} &= \langle P_\mu \cdot P_\nu \rangle -
\langle P_\mu \rangle \langle P_\nu \rangle  \cr
C_{V,D\mu} &= \langle \bar \psi D_0 \psi \cdot P_\mu \rangle -
\langle \bar \psi D_0 \psi \rangle \langle P_\mu \rangle  \cr
C_{V,DD} &= \langle \bar \psi D_0 \psi \cdot \bar \psi D_0 \psi
\rangle -
\langle \bar \psi D_0 \psi \rangle \langle \bar \psi D_0 \psi
\rangle  \cr
}\eqno(20)
$$
Here $\mu,\nu$ denote $\sigma$ or $\tau$. We also left out the
correlations of $P_\mu$ and $ \bar \psi D_0 \psi $
with the chiral condensate $\bar \psi \psi$. These are given already
in eq.~(17) and, due to additional mass factors (see eq.~(16)),
certainly will not lead to a singular behaviour in the specific heat.
From eq.~(5) one expects that the quark mass dependence of the
specific heat peak is controlled by the exponent $z_\alpha = \alpha
/\beta\delta$. In the case of the $O(N)$ models in three dimensions
the situation
becomes, however, a bit more complicated due to the fact that the
exponent $\alpha$ is expected to be negative [4]. In this case
the specific heat is not expected to diverge, but rather should
develop a cusp at $T_c$. Accordingly none of the response
functions defined in eq.~(20) is actually expected to diverge.
They rather should approach a constant in the limit of vanishing
quark mass, with the quark mass dependence described by eq.~(10).

\bigskip \bigskip
\noindent
{\deinfont 4) Numerical Results}
\medskip

We have performed calculations on an $8^3 \times 4$ lattice,
which may
be considered to be quite a small lattice, yet, Ref.~[5] shows
that
varying the spatial lattice extent from $N_\sigma =$ 6 to 12
at $N_\tau = 4$ surprisingly does not change the pseudo-critical
temperature [8]. Likewise, increasing $N_\tau$ from 4 to 8 while
keeping $N_\sigma / N_\tau $ between 2 and 4 also does not lead to
apparent differences in the scaling behaviour of the pseudo-critical
points. Nonetheless, it is important to get control over finite size
effects. The present analysis should  thus be considered as a first
step towards a quantitative determination of critical exponents in QCD
with light fermions.

For the quark masses, we have chosen the values
$m_q = 0.075$, 0.0375 and 0.02 in units of the lattice spacing
$a$.
These quark masses have been selected such that they can provide
additional
information on the quark mass dependence of pseudo-critical
couplings
beyond the existing data [2].
At each value of the quark mass we have performed simulations at
several
values of the gauge coupling, $6/g^2$, in the pseudo-critical
region.
The simulations have been carried out by means of the so-called
hybrid R
algorithm [13]. Guided by analyses of the hybrid
Monte Carlo algorithm, we have scaled the step size according to
the quark mass, $d \tau = 0.05, 0.04$ and 0.03
(in the normalization of [13])
for $m_q = 0.075, 0.0375$ and 0.02 respectively.
The trajectory length was fixed to $\tau = 1$.
For the conjugate gradient inversion we have settled at a
required precision of
$(\Phi - D X)^2 / \Phi^2 \leq 0.5 \times 10^{-13}$.
The autocorrelation times have been determined
to $\tau_{\rm exp} \leq 100$ trajectories right on the
pseudo-critical
couplings, the longest
correlations being in the Polyakov loop, while away from the
cross-over
$\tau_{\rm exp}$ is considerably smaller.
At each value of the coupling we collected between 5000 and 10000
trajectories
which then have been analyzed using a reweighting with the
density of
states at intermediate values of the gauge coupling [14].
Errors were computed by the jack-knife procedure.

\bigskip
\noindent
{\deinfont 4.1) Susceptibilities and the Specific Heat}
\medskip

In Fig.~1 we show the Polyakov loop response function, $\chi_L$,
for three
values of the quark mass. Pronounced peaks are clearly visible.
Although
the location of the peaks clearly depends on $m_q$,
their heights do not have any significant quark mass
dependence. This is
consistent with the analysis presented in Ref.~[8], which also
suggests
that the volume dependence of the peak height is small and
distinctively
different from the situation in the pure gauge theory.

The behaviour of $\chi_L$ is also drastically
different from that of the chiral susceptibility, $\chi_m$, which
is shown in Fig.~2.
The rise in the peak height with decreasing quark mass as well as
the shift in the critical coupling is clearly seen.
In Fig.~2 the
two contributions, $\chi_0$ and $\chi_{\rm conn}$ are shown
separately. The contribution from the connected part
to the susceptibility is a slowly varying function in the
critical region. The ratio $\chi_{\rm conn}/\chi_0$ decreases with
decreasing quark mass. However, even for $m_q=0.02$ we find that
$\chi_{\rm conn}$ contributes about 30\% to
the value of $\chi_{m,{\rm max}}$ and in the region of the zero
quark mass critical temperature it gives the dominant contribution.
This will be
particularly important for our discussion of the chiral cumulant
in section 4.3.

\medskip
$$
\offinterlineskip \tabskip=0pt
\vbox{
\halign to 0.8\hsize
{\strut
 \vrule width0.8pt\quad#
 \tabskip=0pt plus 100pt
 &#\quad
 &\vrule#&
 &\quad #\quad
 &\vrule#
 &\quad #\quad
 &\vrule#
 &\quad #\quad
 &\vrule#
 &\quad #\quad
  \tabskip=0pt
 &\vrule width 0.8pt#\cr
 \noalign{\hrule}\noalign{\hrule}
 & $z_m$ && $z_{t,\sigma}$ && $z_{t,\tau}$ && $z_{t,f}$&& $z_L$ &
\cr
 \noalign{\hrule}\noalign{\hrule}
 &0.79(4)&& 0.63(7) && 0.63(7) && 0.65(7)&& 0.05(6) & \cr
\noalign{\hrule\hrule}}}
$$
\medskip
{\narrower
Table II: Critical exponents controlling the divergence of the
chiral ($m$), Polyakov loop ($L$) and various parts of the thermal
susceptibilities
in the limit of vanishing quark mass. Results are obtained from
straight line fits to the data shown in Fig.~4.
\medskip}

In Fig.~3 we show the partial contributions to the thermal
susceptibility, defined in eq.(17). It is apparent that all three
have a similar quark mass dependence. Note also that the peak
position
in all three cases coincides with the pseudo-critical couplings
in $\chi_L$ or $\chi_m$.

The peak heights of all
susceptibilities are shown in Fig.~4 as a function of quark mass.
A fit with a powerlike singular behaviour as indicated in eq.~(18)
yields the
critical exponents  given in Table II. While the exponent $z_{L}$
is compatible with being zero, we find that $z_{m}$ is in remarkably
good agreement with the $O(N)$ prediction. The agreement is not that good
for the exponent of the thermal susceptibility, which comes out to be
about 50\% larger than expected for an $O(N)$ symmetric model. This,
however, is not that unexpected as the exponent $z_m$ is only sensitive
to the magnetic exponent $y_h$, while $z_t$ is also related to $y_t$.
The latter gives directly the correlation length exponent $y_t=1/3\nu$
and thus is expected to be more sensitive to the spatial size of the system.
We can rephrase our numerical results for $z_m$ and $z_t$ in terms of
the exponents $y_h$ and $y_t$. This yields
$$\eqalign{
y_h &= 0.83 \pm 0.03~~, \cr
y_t &= 0.69 \pm 0.07~~, \cr}
\eqno(21)
$$
which should be compared with the $O(4)$ values $y_h =0.828$ and $y_t
=0.452$.

As has been discussed above the various response functions
contributing to the specific heat are not expected to diverge in
the zero quark mass limit. Results for three of the six different
response functions defined in eq.(20) are shown in Fig.~5 for the three
different quark mass values studied here. A comparison with the
susceptibilities
displayed in Figs.~3 and 4 clearly shows that the increase in the
peak height is considerably slower. However, we do not have any
direct evidence for a finite limit of the peak height in the limit of
vanishing quark mass. Results for the calculated peak values,
$C_{V}^{\rm max}$, and
the corresponding pseudo-critical couplings are given in Table III.
We have analyzed the scaling behaviour of these peaks in two
ways, which clearly show the present uncertainties in the determination of
$z_\alpha$:

\item{(i)}{Assuming that the specific heat diverges in the zero
quark mass limit, \par
one can analyze the quark mass dependence of the specific heat
response functions in the same way as the various components of
the susceptibilities, ie. we fit the peak heights to eq.~(10) with
$c_0\equiv 0$. By construction this leads to $z_\alpha > 0$.
We find, depending on the particular response function, values for
$z_\alpha$ between 0.28 and 0.34. The best fit result yields
$z_\alpha = 0.28(6)$.}

\item{(ii)}{Assuming that the specific heat stays finite in the
zero quark mass limit, \par
one can fit the peak heights to eq.~(21) with
$c_0 > 0$. By construction this leads to $z_\alpha < 0$. As can be
seen in Fig.~5 the strongest quark mass dependence is found in the
purely gluonic response functions. They lead to
$z_\alpha = -0.05(2)$.}

We note that even by assuming a finite specific heat in the limit of
vanishing quark mass our fits suggest a large value of the response
functions in that limit. For instance we find $C_{V,\tau\tau}^{\rm max}
(m_q=0) \simeq 6$.
This indicates that these response function will
still show quite a strong quark mass dependence also for quite small
values of $m_q$, which is related to the nearly vanishing magnitude of
$\alpha$ ($\alpha =0$ corresponds to a logarithmic singularity). The
upper limit for $\alpha$ obtained in this analysis, $\alpha \le 0.34$,
corresponds to an upper limit of 0.6 for the thermal exponent $y_t$. As
can be seen from eq.~(21) this is slightly lower than the result
obtained from the analysis of the thermal susceptibility.

Further calculations at smaller quark masses are thus particularly
interesting for these response functions and are needed in order to
extract the thermal exponent $y_t$ and to further clarify the nature
of the scaling behaviour
of the specific heat in the zero quark mass limit.

\medskip
$$
\offinterlineskip \tabskip=0pt
\vbox{
\halign to 1.0\hsize
{\strut
 \vrule width0.8pt\quad#
 \tabskip=0pt plus 100pt
 &#\quad
 &\vrule#&
 &\quad #\quad
 &\vrule#
 &\quad #\quad
 &\vrule#
 &\quad #\quad
 &\vrule#
 &\quad #\quad
 &\vrule#
 &\quad #\quad
  \tabskip=0pt
 &\vrule width 0.8pt#\cr
 \noalign{\hrule}\noalign{\hrule}
 & ~~ && \multispan9 $C_{V,\mu\nu}^{\rm max}$  & \cr
 & ~~ && \multispan9 $\beta_c$  & \cr
 \noalign{\hrule}\noalign{\hrule}
 & $m$ && $D\sigma$ && $D\tau$ && $\sigma\tau$&& $\sigma\sigma$
&&
$\tau\tau$ & \cr
 \noalign{\hrule}\noalign{\hrule}
 & 0.075 && 0.0535(53) && 0.0691(60) && 0.761(65)&&
0.807(58)&&0.957(71) & \cr
 & ~ && 5.350(4) && 5.350(4) && 5.348(4)&&
5.348(4)&&5.348(4) & \cr
 \noalign{\hrule}\noalign{\hrule}
 & 0.0375 && 0.0722(83) && 0.0921(97) && 0.974(45)&&
0.999(40)&&1.193(51) & \cr
 & ~ && 5.306(4) && 5.306(4) && 5.306(2)&&
5.306(2)&&5.306(2) & \cr
 \noalign{\hrule}\noalign{\hrule}
 & 0.02 && 0.0844(61) && 0.1040(66) && 1.165(62)&&
1.170(53)&&1.409(70) & \cr
 & ~ && 5.282(2) && 5.282(2) && 5.282(2)&&
5.282(2)&&5.282(2) & \cr
\noalign{\hrule\hrule}}}
$$
\medskip
{\narrower
Table III: The peak values of various response functions
contributing
to the specific heat and the corresponding values of the
pseudo-critical
couplings.
\medskip}
\bigskip
\noindent
{\deinfont 4.2) Pseudo-Critical Couplings}
\medskip

The locations of the peaks in the various susceptibilities or the
specific heat, which have been discussed in the previous section,
can be used to define pseudo-critical couplings.
Although these pseudo-critical couplings may differ for different
observables at non-vanishing values of the quark mass, we found
that they agreed within our numerical accuracy. We thus may quote
a common pseudo-critical coupling for all the response functions
analyzed by us. These are collected in Table IV.

It is quite reassuring that the pseudo-critical couplings
extracted from
the location of the peak in $\chi_m$, the various components of
$\chi_t$  and $C_V$ agree this well. In particular this shows that the
pseudo-critical couplings, extracted so far from the variation of the
slope of $\langle \bar\psi \psi \rangle$ as a function of $6/g^2$
(this corresponds to the location of the peak in $\chi_{t,\sigma
(\tau)}$) can indeed be used to discuss the scaling behaviour of these
couplings as a function of the quark mass.

\medskip
$$
\offinterlineskip \tabskip=0pt
\vbox{
\halign to 0.3\hsize
{\strut
 \vrule width0.8pt\quad#
 \tabskip=0pt plus 100pt
 &#\quad
 &\vrule#&
 &\quad #\quad
  \tabskip=0pt
 &\vrule width 0.8pt#\cr
 \noalign{\hrule}\noalign{\hrule}
 & $m_q$ && $6/g_c^2$ & \cr
 \noalign{\hrule}\noalign{\hrule}
 & 0.075 && 5.350(4)  & \cr
 & 0.0375 && 5.306(2) & \cr
 & 0.02 && 5.282(2)  & \cr
\noalign{\hrule\hrule}}}
$$
\medskip
{\narrower
Table IV: Pseudo-critical couplings determined from the location
of the peak in various response functions. Errors
are obtained from a jack-knife analysis of the  interpolation
curves
resulting from a reweighting with the density of states method.
\medskip}

In Fig.~6 we show the quark mass dependence of the pseudo-critical
couplings. Here we also include data from earlier simulations [6,7],
which have been summarized in Ref.~[2]. The pseudo-critical couplings
have been fitted with the ansatz
$$
{6 \over g_c^2(m_q)} = c_0 + c_1 m_q^{z_c}
~~,\eqno(22)
$$
where $c_0$ gives the critical coupling in the zero quark mass
limit and $z_c \equiv 1/\beta\delta$.
We find
$$\eqalign{
{6 \over g_c^2(0)} &= 5.243 \pm 0.010 \cr
{1 \over \beta\delta} &= 0.77 \pm 0.14}
~~.\eqno(23)
$$
Compared to the corresponding $O(4)$ value $1/\beta\delta =0.55(2)$ and
the $O(2)$ value $1/\beta\delta = 0.60(1)$ the three parameter fit
yields a larger value for this combination of exponents. However, also a
two parameter fit with $1/\beta\delta$ fixed to its $O(4)$ value still
yields a good $\chi^2$. We show this fit also in Fig.~6. The resulting
zero quark mass critical coupling is found to be ${6 / g_c^2(0)} =
5.222 \pm 0.003$. We also note that the result obtained
for $1/\beta\delta$ is consistent with the determination of the thermal
and magnetic exponents in the previous section and also suggests that in
our present analysis we overestimate the value for $y_t$.
\bigskip
\noindent
{\deinfont 4.3) The Chiral Cumulant}
\medskip

The most direct procedure to determine the critical exponent $\delta$,
which is independent of any ansatz for a fitting function, is given
through an analysis of the chiral cumulant, defined in eq.~(8).
In the vicinity of the critical point this is a simple scaling
function
and curves for different quark masses cross in a unique point -
the zero mass critical coupling, if
subleading corrections from the non-singular part of thermodynamic
quantities can be ignored. In Fig.~7 we show $\Delta (6/g^2,m_q)$ for
our three different quark masses. For the largest mass the
pseudo-critical region was quite far away from the zero quark
mass critical region and we did not attempt to extend the
calculations for that mass into this regime. The simulations for
the two smaller quark masses, however,
have been extended into this region and we see from Fig.~7 that
they indeed cross at a coupling, which is consistent with the result
obtained
from our fit to the pseudo-critical couplings (Fig.~6). Moreover,
we find that the value of $\Delta$ at the crossing point gives an
astonishingly accurate estimate of $1/\delta$. In fact, due to
the rather weak dependence of $\Delta$ on the coupling $6/g^2$ in
this region, we find that the exponent $1/\delta$ is much better
determined than the zero quark mass critical coupling.
Taking for the latter the interval of values obtained from the two fits
shown in Fig.~6. ie.
$5.22 < 6/g^2_c (0) < 5.25$, we get for the exponent $\delta$,
$$
0.21 < 1/\delta = 0.23 < 0.26
~~.\eqno(24)
$$

\bigskip
\noindent
{\deinfont 5) Conclusions}
\medskip

We have studied various response functions which contribute to
susceptibilities and the specific heat in two-flavour QCD, simulated
with staggered fermions on a lattice of size $8^3\times 4$. We
find that these observables are very well suited to localize the
pseudo-critical couplings at non-vanishing values of the quark
mass. At least at the moderate values of the quark mass analyzed here
the height of the peaks in the susceptibilities is very well determined
and can be used to investigate the scaling behaviour of these quantities.
The analysis of the peak heights in the response functions
as well as the quark mass dependence of the peak location yields
independent observables, which can be used to determine the critical
exponents $\alpha$, $\beta$ and $\delta$. Additional information on
$\delta$ and the location of the zero quark mass critical point is
obtained from the structure of the chiral cumulant. These exponents can
be related to the two basic exponents $y_t$ and $y_h$, which
characterize the scaling behaviour of the singular part of the free
energy density.

Our present analysis yields a magnetic exponent $y_m$, which is
consistent with the value expected for a second order phase transition
controlled by $O(4)$ exponents (eq.~(21)). Our result for the thermal
exponent, however, turns out to be about 50\% larger than the
corresponding $O(4)$ value.

Clearly the errors are still quite large. However, the present results
are consistent with our expectations based on the existence of a second
order phase transition in the limit of vanishing quark mass. We
find it particularly reassuring, that we obtain the weakest quark mass
dependence in response functions which contribute to the specific heat.
We also confirm a weaker quark mass dependence of the
peak of the thermal susceptibility relative to that of the chiral
susceptibility. Moreover, we find that the Polyakov loop response
function is insensitive to changes in the quark mass. This suggests that,
indeed, the chiral symmetry restoration is the driving mechanism for the
finite temperature phase transition in two flavour QCD.

We expect that in particular the thermal exponent $y_t$ is sensitive to
finite lattice effects. The behaviour of the susceptibilities thus
has to be studied for smaller quark masses and on larger lattices.
\vskip 20pt
{\bf Acknowledgements:} This work has been supported in part by the
Deutsche Forsch\-ungsgemeinschaft under contract Pe 340/3-2. The numerical
simulations have been performed on the NEC SX-3 in K\"oln and the
Cray Y-MP of the HLRZ-J\"ulich. We thank the staff of these computer 
centers for their support.

\vfill\eject
\noindent
{\deinfont References}
\medskip
\item{1)} For a recent review see: F. Karsch,
Nucl. Phys. B (Proc. Suppl.) \underbar{34} (1994) 63.
\item{2)} F. Karsch, Phys. Rev. D49 (1994) 3791.
\item{3)} R.D. Pisarski and F. Wilczek, Phys. Rev. \underbar{D29}
(1984) 338.
\item{4)} F. Wilczek, Intern. J. Mod. Phys. \underbar{A7} (1992)
3911;
\item{  } K. Rajagopal and F. Wilczek, Nucl. Phys.
\underbar{B399} (1993) 395.
\item{5)} J.I. Kapusta and A.M. Srivastava, {\it The Proximal Chiral
Phase Transition}, NSF-ITP-94-28.
\item{6)} S. Gottlieb, Phys. Rev. \underbar{D47} (1993) 3619 and
references therein.
\item{7)} F.R. Brown et al., Phys. Rev. Lett. \underbar{65} (1990)
2491. 
\item{8)} M. Fukugita, H. Mino, M. Okawa and A. Ukawa, Phys. Rev.
\underbar{D42} (1990) 2936.
\item{9)} G. Boyd, J. Fingberg, F. Karsch, L. K\"arkk\"ainen and
B. Petersson, Nucl. Phys. \underbar{B376} (1992) 199.
\item{10)} G. Baker, D. Meiron and B. Nickel, Phys. Rev.
\underbar{B17} (1978) 1365.
\item{11)} J.C. Le Guillou and J. Zinn-Justin,
 Phys. Rev. \underbar{B21} (1980) 3976 and J. Phys. Lett. (Paris)
\underbar{46} (1985) L-137.
\item{12)} for a discussion of QCD thermodynamics on the lattice
see for instance:
F.Karsch, Simulating the Quark-Gluon Plasma on the Lattice,
Advanced Series on Directions in High Energy Physics - Vol.6
(1990) 61,
"Quark Gluon Plasma" (Ed. R.C. Hwa), Singapore 1990, World
Scientific.
\item{13)} S. Gottlieb, W. Liu, D. Toussaint, R.L. Renken and
R.L. Sugar, Phys. Rev. \underbar{D35} (1987) 3972.
\item{14)} for a discussion of our implementation of the density
of state
method and references see: J. Fingberg, U. Heller and F. Karsch,
Nucl. Phys. \underbar{B392} (1993) 493.

\vfill\eject
\noindent
{\deinfont Figure Captions}
\medskip
\itemitem{Figure 1:} The Polyakov loop response function versus $6/g^2$
for three values of the quark mass. Shown are results from a reweighting
analysis with bins of length $\Delta(6/g^2)=0.002$ close to the
pseudo-critical coupling and 0.005 elsewhere. The couplings at which
simulations have actually been performed are marked by filled symbols.
\medskip
\itemitem{Figure 2:} The chiral susceptibility versus $6/g^2$
for three values of the quark mass. Results from a reweighting
analysis for the disconnected part $\chi_0$ and the connected part
$\chi_{\rm conn}$ defined in eq.~(13) are shown separately.
\medskip
\itemitem{Figure 3:} The three response functions which contribute to
the thermal susceptibility defined in eqs.~(15)-(17).
\medskip
\itemitem{Figure 4:} Peak values of the Polyakov loop response function
($\chi_L$), the chiral ($\chi_m$) and the thermal ($\chi_t$) susceptibility
versus the quark mass.
\medskip
\itemitem{Figure 5:} Three response functions which
contribute to the specific heat as defined in eq.~(20).
\medskip
\itemitem{Figure 6:} Pseudo-critical couplings, $6/g^2_c(m_q)$,
versus $h=N_\tau m_q$ for
$N_\tau =4$. The three new values analyzed here are shown as full
diamonds. The other data points are taken from the collection given in
Ref.~[2]. The dashed curve is a three parameter fit with eq.~(22). The
solid curve is a two-parameter with this function and $z_c$ fixed
to the $O(4)$ value $z_c=0.55$.
\medskip
\itemitem{Figure 7:} The chiral cumulant, $\Delta$, versus $6/g^2$
for three values of the quark mass.

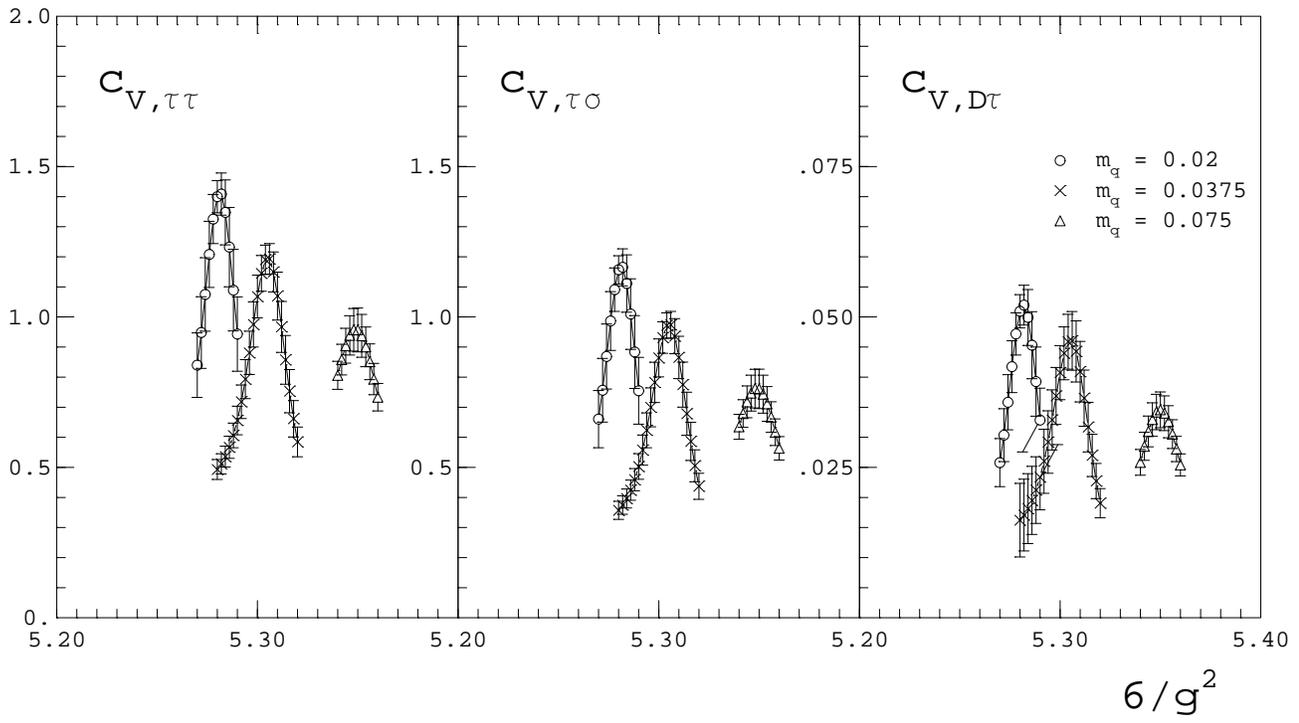
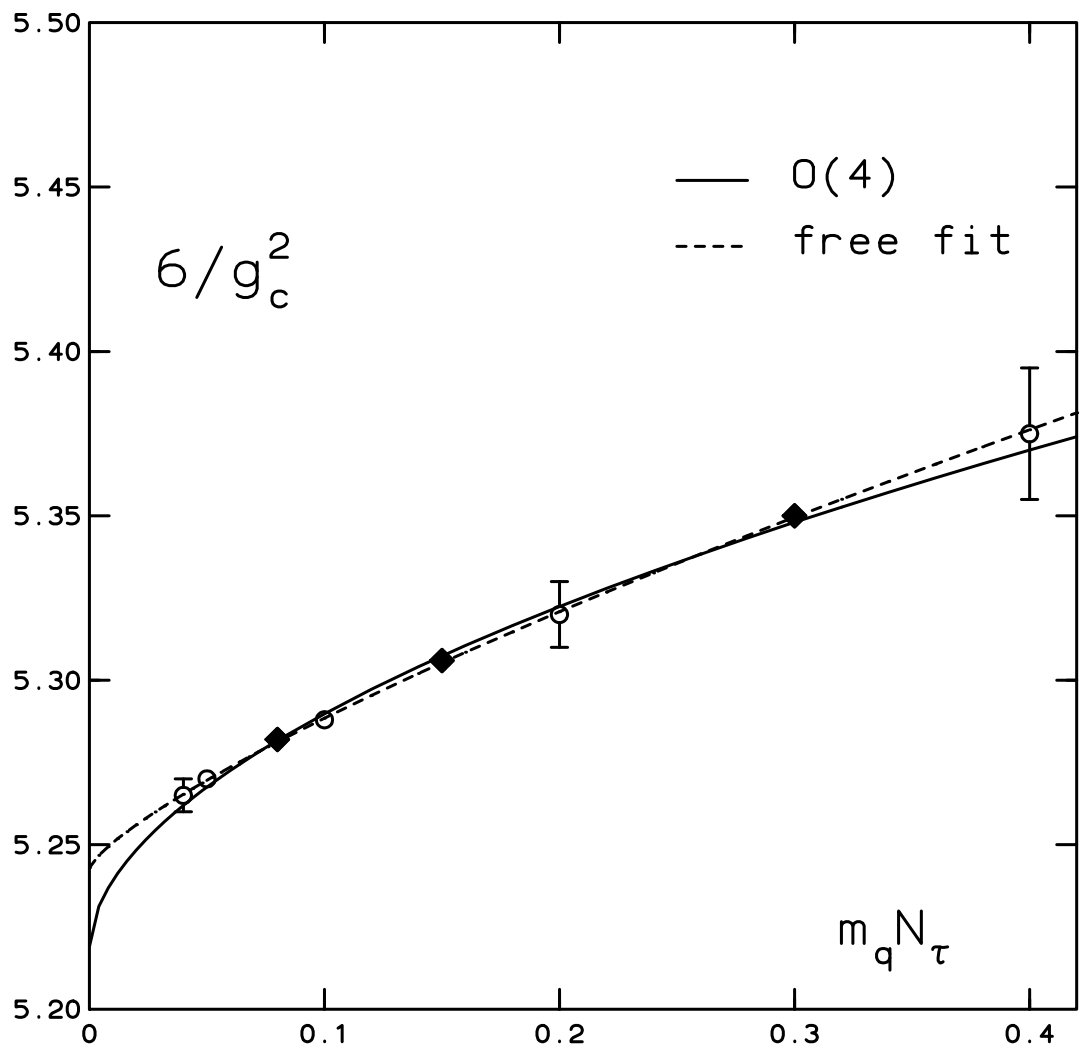
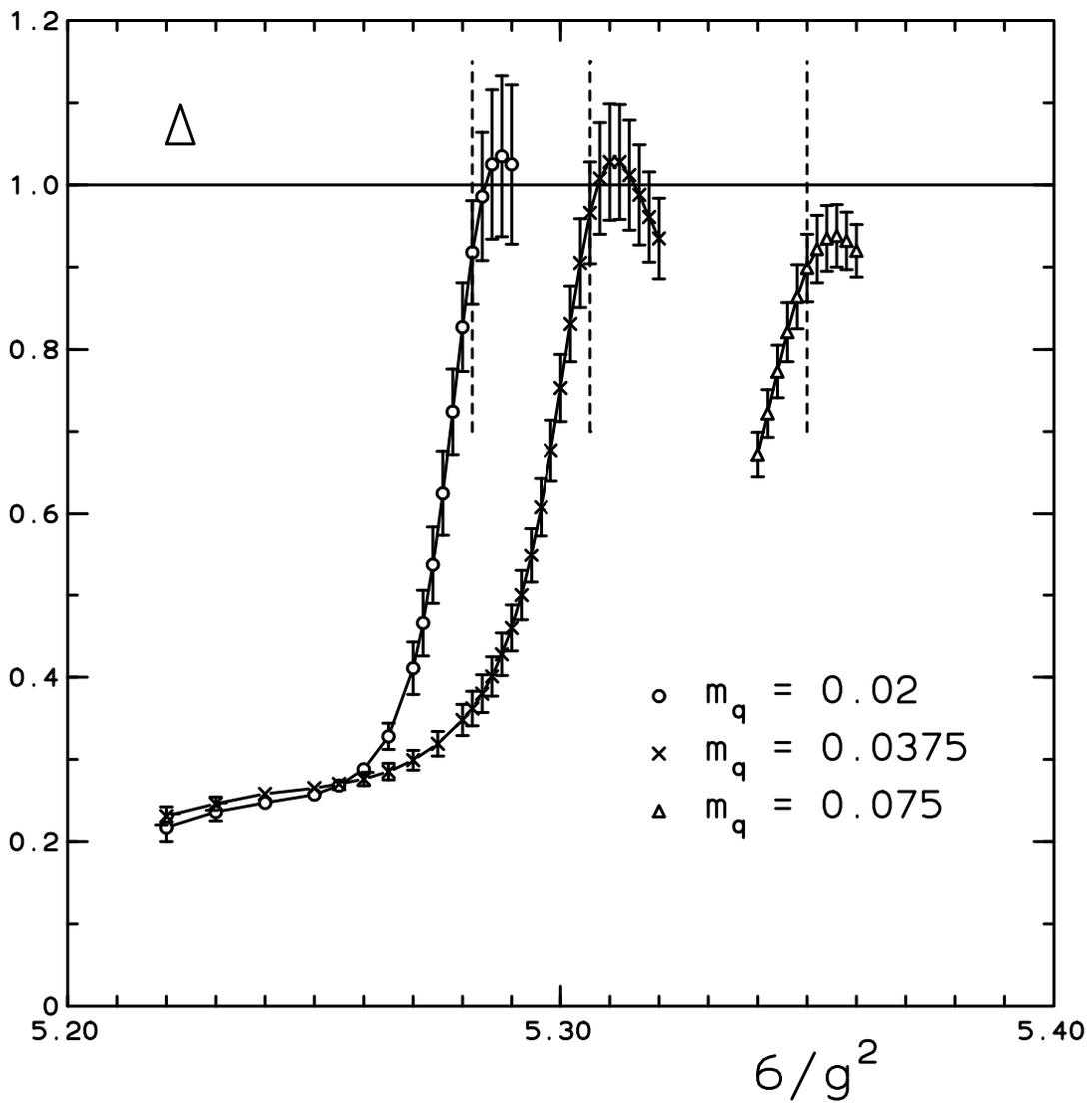
\end